\documentclass{aa}

\usepackage{newtxtext,newtxmath}
\usepackage{amsmath,ulem}
\usepackage{physics}
\usepackage{multirow}
\usepackage{caption}
\usepackage{tablefootnote}

\usepackage[T1]{fontenc}
\DeclareRobustCommand{\VAN}[3]{#2}
\let\VANthebibliography\thebibliography
\def\thebibliography{\DeclareRobustCommand{\VAN}[3]{##3}\VANthebibliography}

\usepackage{dblfloatfix}
\usepackage{graphicx}	
\usepackage{amsmath}	
\usepackage{amssymb}	
\usepackage{hyperref}
\usepackage{xcolor}
\usepackage{float}

\title{Interaction between disk and extended corona in a general relativistic framework}

\subtitle{I. Static slab corona in global energy balance with the underlying disk}

\author{
Sudeb Ranjan Datta \inst{1,2},
Michal Bursa \inst{1}, Michal Dovciak \inst{1}, Wenda Zhang \inst{3} 
}

\institute{
Astronomical Institute of the Czech Academy of Sciences, Bo\v{c}n\'\i -II 1401, Praha 4, Prague, 141~00, Czech Republic\\
\email{datta@asu.cas.cz}
\and
Inter-University Centre for Astronomy and Astrophysics, Post Bag 4, Ganeshkhind, Pune - 411007, India
\\
\and
National Astronomical Observatories, Chinese Academy of Sciences, 20A Datun Road, Beijing 100101, China
}

\titlerunning{Disk corona interaction}  
\authorrunning{Datta et al.}

\date{Received September 15, 1996; accepted March 16, 1997}
\begin{document}

\abstract
{The energy equilibrium between the corona and the underlying disk in a two-phase accretion flow sets a lower limit on the achievable photon index. A slab coronal geometry may not adequately explain the hard state observations of X-ray binaries (XRBs).} 
{We incorporated energy feedback to the accretion disk resulting from illumination by an extended corona, and vice versa. The interaction between these two components makes it possible to find an energetically self-consistent equilibrium solution for a given disk–corona system.}
{We upgraded the existing Monte Carlo radiative transfer code, MONK, to incorporate the interaction between the disk and the extended corona within the general relativistic framework. We introduced an albedo parameter to specify the fraction of the incident flux that is reflected by the disk, while the remainder is absorbed and added to the intrinsic dissipation. Reflection was modeled assuming a semi-infinite electron atmosphere. For a given disk–corona system, Comptonization by the corona and disk illumination are iteratively computed to reach equilibrium, under the constraint that the total observed luminosity equals the total available accretion power.}
{We find global equilibrium solutions not only for the hard state but also for intermediate states, with partial contributions from the disk and from the corona. A higher black hole spin, higher coronal temperature, and higher albedo all lead to harder spectra. For typical coronal temperatures and disk albedos, the lowest achievable photon index with a static slab corona fully covering the disk is approximately 1.7–1.8. Under the assumption of a spatially uniform, extended slab corona, energy must flow from the inner to the outer annuli if local equilibrium is imposed between the disk and the corona at each radial annulus, in addition to the global equilibrium condition.}
{With the upgraded version of MONK, we are now able to achieve global energy equilibrium for a given disk–corona system. This approach holds significant potential for constraining the coronal geometry using not only the observed flux, but also polarization. A static slab does not appear to be a favorable coronal geometry for the hard state of X-ray binaries, even when global energy balance between the disk and corona is taken into account. In future work, we will explore truncated disk geometries and outflowing coronae as potential alternatives.}


\maketitle

\section{Introduction} 
\label{sec_intro}
The first successful model describing angular momentum transport and the formation of an accretion disk, and produced multi-temperature blackbody radiation from low-mass X-ray binaries (XRBs), was formulated by \citet{Pringle1972} and \citet{Shakura1973}. This model was soon extended into a general relativistic framework \citep{Novikov1973, Page1974}, which remains the most widely used model for spectral modeling of accretion disks to this day (see \citealt{Frank2002book, Kato2008book} for further details). However, this ``cold disk model'' cannot account for the higher-energy spectral data ($\gtrsim 10$ keV) observed in XRBs \citep{Lightman1975}. To explain the origin of hard X-rays, incomplete thermalization in an optically thin hot flow must be invoked \citep{Shapiro1976}. To render this hot flow thermally stable, some degree of advection is required \citep{Abramowicz1995, Chakrabarti1995, Narayan1995}. In the unified picture of accretion flows, a truncated disk geometry is therefore proposed, where an optically thick, geometrically thin disk is truncated at a radius inside of which a hot, advective, optically thin, and geometrically thick flow exists. The truncation radius moves inward with increasing accretion rate \citep{Esin1997}. The hot electron medium responsible for producing a hard power law spectrum via inverse Comptonization is referred to as the corona. This truncated geometry is supported by numerous observational studies, both spectral and timing-based \citep{Basak2016, Marcel2019, Zdziarski2020, Marino2021, Banerjee2024, Chand2024, DeMarco2021, Kawamura2022}.

Although truncated disk geometry remains a plausible configuration for accretion flows, the observation of broad fluorescent iron lines from XRBs, and from active galactic nuclei (AGNs), suggests that the optically thick disk often extends close to the black hole (BH) and is illuminated by a hard X-ray–emitting corona even in the hard state \citep{Fabian1989, Ross1996, Miller2006} (for more recent observations, see \citet{WangJi2018, Buisson2019, Chakraborty2020, Sridhar2020}; for AGNs, see \citet{Tanaka1995, Nandra1997, Reynolds1997, Fabian2000}). If the disk extends close to the BH, then the corona must lie above the disk. Buoyant magnetic flares and an optically thin disk atmosphere produced via disk evaporation could also serve as a hot corona \citep{Meyer2000}. However, whether the inner region of the optically thick disk can evaporate completely, thus giving rise to the truncated disk scenario, remains an open area of research \citep{Rozanska2000, Liu2002, Cho2022, Bambic2024}. Other geometric signatures, such as steep emissivity profiles or the reprocessed blackbody component from reflection, have not proven sufficient to distinguish among competing geometries \citep{Svoboda2012, Datta2024}. Recent polarization measurements in the hard state of XRBs suggest that the corona is extended horizontally, perpendicular to the disk axis (\citealt{Krawczynski2022PolCygX1, Ingram2024, Podgorny2024}; for a review, see \citealt{Dovciak2024}). In this work, we also adopt the assumption of a two-phase accretion flow, consisting of a hot, extended slab corona situated above the disk.

For typical parameters of a standard Shakura–Sunyaev disk in X-ray binaries (XRBs), the thermal and viscous timescales are on the order of minutes and days, respectively \citep{Pringle1981}. These are much longer than the light travel time between the disk and the corona, which is approximately $GM/c^3 \sim 10^{-5}$ seconds (where $G$ is the gravitational constant, $M$ is the black hole mass, and $c$ is the speed of light). Moreover, the typical timescale for XRB observations is $\gtrsim 10$ ks. Therefore, in a two-phase accretion flow, we expect the disk and corona to achieve equilibrium through mutual interaction.

For this work we considered the disk–corona connection solely through two processes: the supply of seed photons from the disk for Comptonization, and illumination of the disk by the corona, an approach primarily motivated by observational evidence. The first modeling of such a two-phase accretion flow in mutual equilibrium was presented by \citet{Haardt1991, Haardt1993} (hereafter HM93), and was later developed further by \citet{Stern1995}, \citet{Poutanen1996}, and \citet{Malzac2001} (with a dynamic corona). \citet{Malzac2005} extended this by treating the disk as a slab with either constant density or constant pressure. The most recent advancement in this direction is by \citet{Poutanen2018}, who incorporated reflection spectra using the XILLVER tables \citep{Garcia2010, Garcia2013} to more accurately model reprocessing from cold media.

A key result emerging from these models is that, for a given coronal temperature and geometry, there exists an upper limit to the Comptonized power the corona can produce in equilibrium. Beyond this limit, no energy-balanced solution is possible, even if the entire accretion power is assumed to be dissipated in the corona. This implies that one cannot arbitrarily increase the optical depth at fixed coronal temperature to obtain a harder Comptonized spectrum. As the optical depth increases, the coronal output increases, which enhances disk illumination and, in turn, raises the seed photon flux for Comptonization. For a sufficiently hot and optically thick corona, this feedback creates a runaway cascade, preventing equilibrium from being established. This limiting behavior offers an important means of constraining the disk–corona geometry. The general conclusion is that the hardest observed spectra in XRBs cannot be explained by a static slab corona located above the disk. Viable alternatives include an outflowing corona \citep{Malzac2001} or a truncated disk \citep{Poutanen2018}. In the truncated disk scenario, a sufficient supply of seed photons for inverse Comptonization must still be maintained, either via overlap between the disk and corona or through non-thermal processes such as bremsstrahlung or synchrotron emission originating within the corona itself.

All of the aforementioned studies assume a local energy balance, i.e., they do not consider radial variations in disk emission or feedback from the corona. Moreover, general relativistic effects are typically neglected, even though they are expected to be significant in the vicinity of the black hole. On a different front, recent developments such as the Monte Carlo radiative transfer code MONK \citep{Zhang2019} allow the computation of spectra from disk–corona systems around black holes within a fully general relativistic framework. This code can accommodate a range of extended coronal geometries, including slab, spherical, wedge, and conical structures, both static and outflowing. However, it does not incorporate feedback from the corona to the disk, making it inconsistent with the expected light travel timescale between the two regions and limiting its self-consistency. 

For this work we further developed the MONK code to incorporate the mutual interaction between the disk and the corona. This extension allowed us to find equilibrium solutions for a given black hole, disk, and coronal configuration, thereby establishing a robust framework for constraining system parameters from observations. A key distinction from previous studies lies in our treatment of the fraction of accretion energy dissipated in the corona. Earlier works generally assumed that all of the accretion energy is dissipated in the corona, and that the seed photons for Comptonization arise solely from the reprocessing of coronal illumination by the cold disk (see also Section 2.4 of \citealt{Poutanen2018}, where a small intrinsic disk dissipation component was introduced as a free parameter to achieve equilibrium). This approach was primarily aimed at modeling the hardest spectra achievable and demonstrating the limitations of static slab corona in explaining the hard state of accreting systems. Under that assumption, the equilibrium coronal temperature was derived for a given optical depth. In contrast, we compute the equilibrium solution for given disk and coronal properties by adjusting the fraction of accretion power dissipated intrinsically in the disk through viscous processes, parameterized by $\alpha$, for a fixed coronal electron temperature and optical depth. This is done under the constraint that the total luminosity from the disk–corona system equals the total gravitational power available through accretion. This approach allows us to determine the physically consistent coronal luminosity fraction, not only for the hard state, but also for intermediate states, where a significant contribution from the intrinsic disk blackbody emission may be present.

We describe the complete structure of MONK step by step to find the equilibrium disk-corona solution in Sect. \ref{sec_model}. In Sect. \ref{sec_results} we show the effects of imposing equilibrium on different spectral components and the limiting behavior on the hardest spectra possible to achieve. The effect of BH spin, coronal temperature, and disk albedo are studied. Finally after discussing a few crucial points and our limitations in Sect. \ref{sec_discussion}, we conclude in Sect. \ref{sec_conclusion}.

\section{The complete model}
\label{sec_model}
The Monte Carlo radiative transfer code MONK \citep{Zhang2019} computes Comptonized spectra in Kerr space-time, providing the emergent spectrum from a disk–corona system consisting of a disk blackbody and a Comptonized power law from an extended corona. To generate the final spectrum, the disk and coronal properties must be specified. For the disk, MONK assumes a thin Novikov–Thorne (NT) profile \citep{Novikov1973, Page1974}, requiring input parameters such as the black hole mass and spin, the mass accretion rate, and the radial extent of the disk. These determine the blackbody photon distribution emitted from different annuli. For the corona, one must provide the optical depth, electron temperature, and geometric configuration. Although MONK accurately models disk–corona spectra in a fully relativistic framework, it does not currently incorporate any feedback from the corona to the disk. However, based on earlier studies of two-phase accretion flows, a self-consistent treatment of disk and corona emission is expected to emerge naturally when such feedback is included. This is the gap our work aims to address.

\subsection{Steps in MONK}
\label{sec_procedure_steps}
Here we briefly describe the procedure by which MONK computes the final spectra, as well as the iteration scheme used to achieve equilibrium between the disk and corona. The steps are as follows:

(i) For each radial annulus of the disk, null geodesics are calculated in Kerr geometry to track photon propagation. A photon emitted from the disk can follow one of four possible trajectories: reaching to the observer at infinity, reaching to the corona, hitting back to the disk (self-irradiation) and lost within the horizon of the BH. 

(ii) Photons reaching the corona undergo Compton scattering with hot electrons, governed by the Klein–Nishina cross section. The number of scatterings and the associated energy exchange are determined by the specified coronal properties. For further details on steps (i) and (ii), see \citet{Zhang2019}. The subsequent steps describe the upgrades to MONK implemented in this work to incorporate disk–corona interactions and to achieve an equilibrium solution via iterative procedures.

(iii) Comptonized photons from the corona can either escape to infinity, illuminate the disk, or fall into the black hole. An albedo parameter is introduced to specify the fraction of the illuminating flux that is reflected by the disk; the remaining fraction is absorbed and added to the intrinsic disk dissipation. Reflection is modeled using Chandrasekhar’s multiple scattering formalism, assuming the disk acts as a semi-infinite electron atmosphere.

(iv) As are coronal photons, the self-irradiated photons are reflected and/or absorbed by the disk depending on the albedo. 

(v) The reflected photons have the same four possibilities as the original blackbody photons. If they reach to the corona, they are further Comptonized. We note that here we ignore the second and higher order reflection or absorption of the photons after first reflection.

(vi) All the above steps lead to the total luminosity from the disk-corona system L$_{\rm tot}$ considering the photons available at infinity and the photons lost in the BH, 
\begin{multline}
    L_{\rm tot}=L_{\rm disk, inf}+L_{\rm corona, inf}+L_{\rm refl-selfirr, inf}\\
    +L_{\rm refl-corona, inf}+L_{\rm disk, BH}+L_{\rm corona, BH}.
\label{eqn_total_luminosity_monk}
\end{multline}
Different luminosity components are categorized depending on the different possible sources and destinations of the photons. The first word in the subscript (i.e. disk, corona, refl-selfirr, refl-corona) represents the source of the photon, whereas the last word (i.e., inf, BH) represents the final destination of the photon. For example, $L_{\rm refl-selfirr, inf}$ is the luminosity reaching to infinity from the reflection of self-irradiated photons, and $L_{\rm corona, BH}$ is the luminosity lost due to coronal photons ending in the BH.

(vii) Absorption of the illuminating photons by the disk increases its effective temperature, thereby modifying the disk blackbody photon distribution that serves as the primary source of seed photons for Comptonization. This, in turn, alters the coronal luminosity. Additionally, Comptonization of reflected photons further modifies the total luminosity, $L_{\rm tot}$. Therefore, we iterate the above steps until $L_{\rm tot}$ converges, i.e., changes by less than 1\% between two consecutive iterations. This converged value represents the total luminosity of the disk–corona system in equilibrium. Hereafter, whenever we refer to $L_{\rm tot}$, we mean this final, converged luminosity.

(viii) If the coronal electrons are hotter compared to the seed blackbody photons on average, the corona contributes to the luminosity significantly and $L_{\rm tot}$ becomes larger than the total accretion power available,
\begin{equation}
    L_{\rm acc}=\eta\dot{M}c^2,
\label{eqn_L_acc}
\end{equation}
where $\eta$ is the accretion efficiency governed by the spin of the BH, $\dot{M}$ is the mass accretion rate. If this situation lasts a while, the accretion energy should be dissipated in the corona to keep the electrons hot, otherwise the corona will cool down almost instantly. Therefore, to achieve the balance, we introduce a fraction $\alpha$, which splits the viscous power generated by accretion into part that is radiated away and part $(1-\alpha)L_{\rm acc}$ that is used to energize the corona ($L_{\rm intr}$ and $L_{\rm sc}$ in Appendix \ref{appendix_theoretical_estimates}). With this modification, we note that the total available accretion power remains the same ($L_{\rm acc}$ in equation \ref{eqn_L_acc}); the factor $\alpha$ only scales the intrinsic disk dissipation and consequently the disk blackbody photons distribution. Therefore the fundamental condition for the iteration of $\alpha$ becomes $L_{\rm tot}=L_{\rm acc}$. We make no assumption about the mechanism that transfers ($1\!-\!\alpha$) fraction of the disk viscous energy to the corona to heat up electrons.

(ix) Typically, we start the computation with $\alpha=1$. $\alpha$ remains fixed until convergence is achieved in $L_{\rm tot}$ through iterating the MONK computations for the given input parameters. After reaching to the converged $L_{\rm tot}$ we modify the $\alpha$ so that
\begin{equation}
    \alpha_{\rm new}=\frac{L_{\rm acc}}{L_{\rm tot}}\alpha_{\rm old},
\label{eqn_alpha_iteration}
\end{equation}
where $\alpha_{\rm new}$ is the value for the next run and $\alpha_{\rm old}$ is the value from the last run. Once $\alpha$ is modified, it changes the seed photon distribution for Comptonization and as a consequence will change the $L_{\rm tot}$. Therefore, we again iterate to reach the convergence of $L_{\rm tot}$ keeping $\alpha$ the same. And the cycle repeats. Basically iteration of $\alpha$ is done to reach to the equilibrium solution where each iteration step consists of iterations on $L_{\rm tot}$. The nonsimultaneity of iterations for $\alpha$ and $L_{\rm tot}$ keeps the iteration scheme stable. Equation (\ref{eqn_alpha_iteration}) ensures that as the $L_{\rm tot}$ approaches to $L_{\rm acc}$, the modification factor for $\alpha$ approaches to unity.

(x) Finally, we terminate the computation when convergence is achieved in both $L_{\rm tot}$ and $\alpha$ simultaneously—that is, when the changes in both quantities lie within 1\% between two consecutive iterations. This ensures that the total luminosity from the disk–corona system equals the available accretion power, $L_{\rm acc}$, with a fraction $\alpha$ (uniform across all radii) dissipated in the disk and the remaining fraction dissipated in the corona. The value of $\alpha$ is determined self-consistently and depends on the specified disk and coronal properties.

(xi) It is important to note that the above method does not always guarantee a steady-state equilibrium solution. If the corona is too hot and/or too optically thick, the feedback loop between the disk and corona becomes unstable, and no self-consistent equilibrium can be reached. In such cases, the computation is terminated once the iterated value of $\alpha$ drops below $10^{-10}$.

The luminosity computation is relatively insensitive to the grid size and the number of photons injected per geodesic. Therefore, we perform the iterative steps using a low-resolution setup, which significantly reduces computational time and resource usage. Once the equilibrium is achieved, we increase the resolution in the final step to obtain the detailed spectra.

\subsection{Modeling the reflection}
\label{sec_model_reflection}
MONK uses a superphoton scheme to do the whole computation (see Sect. 2 in \citealt{Zhang2019} for more details). Each superphoton represents a bundle of identical photons which is parameterized by: energy at infinity ($E_\infty$), weight (w), polarization degree ($\delta$), four-position ($x^\mu$), wavevector ($k^\mu$), and polarization vector ($f^\mu$). The weight (w) signifies the photon generation rate per unit time in a distant observer's frame. Therefore, in MONK, when we say photon, we basically mean a superphoton. The reflection from the semi-infinite electron atmosphere of the disk is diffuse in nature which is spanned in all directions above the disk. Therefore, for each incident photon one reflected photon is produced in a direction chosen randomly from uniform distribution of cos$\theta$ (within 0 and 1) and $\phi$ (within 0 and 2$\pi$) where $\theta$ and $\phi$ are the polar and azimuthal angle in spherical polar coordinates in the local rest frame above the disk. The energy of the reflected photon becomes 
\begin{equation}
    E_{\rm refl} = \frac{p_\mu^{\rm inc}u^\mu}{p_\mu^{\rm refl}u^\mu}E_{\rm inc}.
\end{equation}

\begin{figure*}
\centering\includegraphics[width=\textwidth]{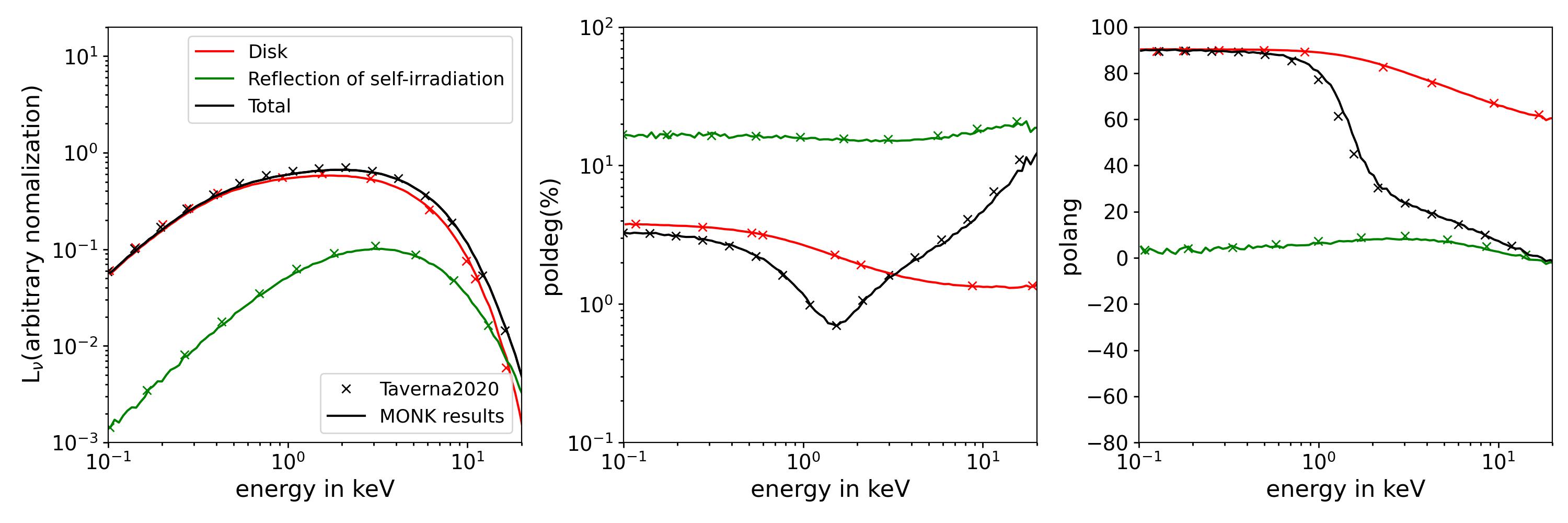}
\caption{Benchmarking the self-irradiated reflection computed using MONK, assuming the disk to be a semi-infinite electron atmosphere, against the results of \citet{Taverna2020} for a Kerr black hole with spin parameter $a = 0.998$ and an inclination angle of 75$^\circ$. The data points from \citet{Taverna2020} were extracted using the \href{https://automeris.io/}{\texttt{WebPlotDigitizer}}}
\label{fig_benchmark_Taverna}
\end{figure*}

Here, $E_{\rm refl}$ and $E_{\rm inc}$ denote the energies of the reflected and incident photons, respectively, as measured by an observer at infinity. The four-momenta of the incident and reflected photons in Boyer-Lindquist coordinates are represented by $p_\mu^{\rm inc}$ and $p_\mu^{\rm refl}$, and $u^\mu$ is the four-velocity of the disk particle at the location where the incident photon strikes the disk. The weight of the reflected photon is calculated using Chandrasekhar's multiple scattering formula for a semi-infinite electron atmosphere (equation 163, Section 70.3, Chapter X in \citealt{Chandrasekhar1950}). This scattering matrix depends on the polar angles of the incident and reflected directions, as well as the difference in their azimuthal angles. Chandrasekhar’s formulation ensures conservation of total energy flux between the incident and reflected radiation. To incorporate partial reflection, we multiply the computed weight by the disk albedo (ranging from 0 to 1), which specifies the fraction of incident flux that is reflected. The remaining fraction $(1 - \text{albedo})$ is absorbed and added to the intrinsic viscous dissipation of the disk. In addition to determining the weight, Chandrasekhar’s scattering matrix also provides the polarization of the reflected photon. We validate our implementation of self-irradiation reflection by benchmarking it against the results of \cite{Taverna2020}, which in turn was benchmarked against the earlier work of \cite{Schnittman2009}. A comparison for a spin parameter ${\rm a} = 0.998$ and an inclination of 75$^\circ$ is presented in Fig.~\ref{fig_benchmark_Taverna}. 

To improve photon statistics at high energies, photons are split into multiple photons with appropriately scaled weights during scattering in the optically thin corona \citep{Dolence2009, Zhang2019}. These higher-energy photons then illuminate the disk and contribute to the coronal reflection component. Through iterative coupling between the disk and the corona, we also account for the Comptonization of reflected photons that re-enter the corona. However, to prevent an exponential increase in the total number of photons during successive iterations, photon splitting during Comptonization is applied only to the seed blackbody photons emitted from the disk.

\subsection{Absorption of illumination}
The value of the albedo depends on the ionization state of the disk. Even for highly ionized disks, a significant portion of the incident radiation is absorbed and reprocessed into lower-energy emission \citep{Ross2005, Garcia2013}. Typically, the albedo is expected to be 0.5 or lower. Consequently, absorption of the illuminating flux plays a crucial role in heating the disk and maintaining the overall energy balance between the disk and the corona. To compute the illuminated flux per unit area in the local corotating frame of the disk, we transform the area element and proper time from the distant observer’s frame into the local frame, following the same prescription as in Section 2.1 of \cite{Zhang2024}. This procedure is effectively the inverse of that used for calculating blackbody photon emission from the disk and transforming it to the observer’s frame, as described in \cite{Zhang2019}. The illumination for each radial annulus is computed by including contributions from both the corona and self-irradiation. As noted in Section \ref{sec_procedure_steps}, we scale the effective mass accretion rate, and hence the energy flux from intrinsic disk dissipation, by the parameter $\alpha$. Therefore, the effective temperature of the NT disk at a radial distance $r$ from the BH is
\begin{equation}
    \sigma T_{\rm eff}^4(r)=\alpha F_{\rm NT}(r)+(1-{\rm albedo})F_{\rm ill}(r).
    \label{eqn_Teff}
\end{equation}
Here, $F_{\rm NT}(r)$ denotes the flux due to intrinsic viscous dissipation at radius $r$, determined by the black hole mass and mass accretion rate following \cite{Novikov1973}, while $F_{\rm ill}(r)$ is the flux incident on that annulus from illumination. We assume that both the energy dissipation fraction $\alpha$ and the albedo are uniform across the disk. After each iteration of $\alpha$, seed blackbody photons are generated using the updated effective temperature, following the standard prescription in MONK. The disk atmosphere is modeled as a semi-infinite planar atmosphere \citep{Chandrasekhar1950}, which MONK uses to compute the polarization properties and angular distribution of blackbody emission (see Section 2.5 of \citealt{Zhang2019} for details). From equation \ref{eqn_Teff}, it is evident that even if the intrinsic dissipation component decreases due to a reduction in $\alpha$ (as required for equilibrium in a given disk-corona system), the total effective temperature—and thus the disk's blackbody emission—can still remain significant, depending on the amount of illumination and the assumed albedo.

\subsection{Photon index of the Comptonized radiation}
\label{sec_fitting_power_law}
In MONK, the corona is characterized by its geometry, optical depth, and electron temperature. However, from an observational standpoint, the photon index ($\Gamma$) of the Comptonized power law component is a more directly relevant quantity. To extract this, we fit the Comptonized photon flux reaching infinity using a simple power law model, $N(E) = N_0,E^{-\Gamma}$, where $N(E)$ is the photon flux per unit energy. The fitting is performed using the \href{https://docs.scipy.org/doc/scipy/reference/generated/scipy.optimize.curve\_fit.html}{\texttt{curve\_fit}} function from SciPy. A nontrivial aspect of this procedure is the determination of the appropriate energy range for fitting, which can vary with the input parameters. Since the power law shape emerges from the superposition of contributions from different orders of Compton scatterings \citep{Rybicki1979}, the fitting range must lie between the average energy of singly scattered photons and that of the highest significant scattering order. This ensures that we capture the regime where multiple orders of scattering overlap to form a power law. MONK provides information about the scattering order and energy distribution, which we use to define this fitting range. Given the low optical depths considered ($\tau < 1$), the number of photons decreases rapidly with increasing scattering order, making the statistics poor at higher orders. To avoid this, we limit the analysis to those scattering orders that account for at least 0.1\% of the total number of scattered photons. This criterion ensures reliable statistical sampling without losing physically important contributions. With increasing optical depth and keeping rest of the parameters the same, the energy range for fitting becomes broader and broader. After finding the energy range, we interpolate the scattered radiation of the MONK output in linear energy grid with bin size of 0.25 keV to do the fitting. We note that the estimation of the energy range for fitting does not depend on the inclination. After finding the energy range, we fit the scattered photons at different inclinations to infer about the photon index.

\section{Results}
\label{sec_results}
Although MONK is capable of performing computations for a wide range of extended coronal geometries—such as spherical, slab, conical, wedge, outflowing slab, and truncated configurations—here we focus solely on the classical case of a two-phase accretion flow \citep{Haardt1991}: a static, thin slab corona that fully covers the accretion disk and corotates with it. The primary motivation is to benchmark the current developments and facilitate direct comparison with results from existing literature. In a follow-up study, we will extend this analysis to explore the effects of alternative coronal geometries.

\subsection{Reference values of input parameters in MONK}
\label{sec_input_parameters}

As reference values of the input parameters, we assume the mass of the BH, $M=10M_\odot$, spin, ${\rm a}=0$, color correction factor, $f_{\rm col}=1.8$, ${\rm albedo}=0.5$, $\dot{M}=0.1\dot{M}_{\rm Edd}$ ($\dot{M}_{\rm Edd}\!=\!2.45\times10^{18}$ g/s) where $\dot{M}_{\rm Edd}=L_{\rm Edd}/\eta c^2$, $L_{\rm Edd}$ is the Eddington luminosity. The equatorial razor thin NT disk is extended from the innermost stable circular orbit (ISCO) to 500 $R_{\rm g}$. The slab corona, corotating with the Keplerian disk, has vertical extension from 0 to 0.1 $R_{\rm g}$ and lie above the disk with the same radial extension as the disk. We fix the electron temperature of the corona to be 120 keV (higher boundary of the electron temperature typically observed in hard state, \citealt{Poutanen2018}). In addition to the above input parameters, we need to fix the optical depth of the corona to find the equilibrium solution. In Sect. \ref{sec_effect_of_equilibrium} we fix it to a specific value, whereas in later sections we find the equilibrium solutions with increasing optical depths till we reach $\alpha\sim0$. In Sect. \ref{sec_effect_spin_Te_albedo} when we study the effect of different input parameters, except the investigated parameter, we fix all the other input parameters to their reference values.

\subsection{Equilibrium solution with global energy balance}

Here we present the results of the equilibrium solutions for a given disk–corona system. Equilibrium is achieved by iteratively adjusting the disk power fraction $\alpha$ that splits the viscous power between the disk and corona as described in Section \ref{sec_model}. The fundamental condition for reaching equilibrium is that the total radiative luminosity from the system matches the total available accretion power, i.e., $L_{\rm tot} = L_{\rm acc}$. Unless stated otherwise, all results presented below correspond to the final equilibrium solutions obtained after iterating over $\alpha$. The only exception is the $\alpha = 1$ case discussed in Section \ref{sec_effect_of_equilibrium}, shown as the dashed line in Fig.~\ref{fig_fraction1_consistent}.

\subsubsection{Change in spectra due to equilibrium}
\label{sec_effect_of_equilibrium}

\begin{figure}
\centering\includegraphics[width=\columnwidth]{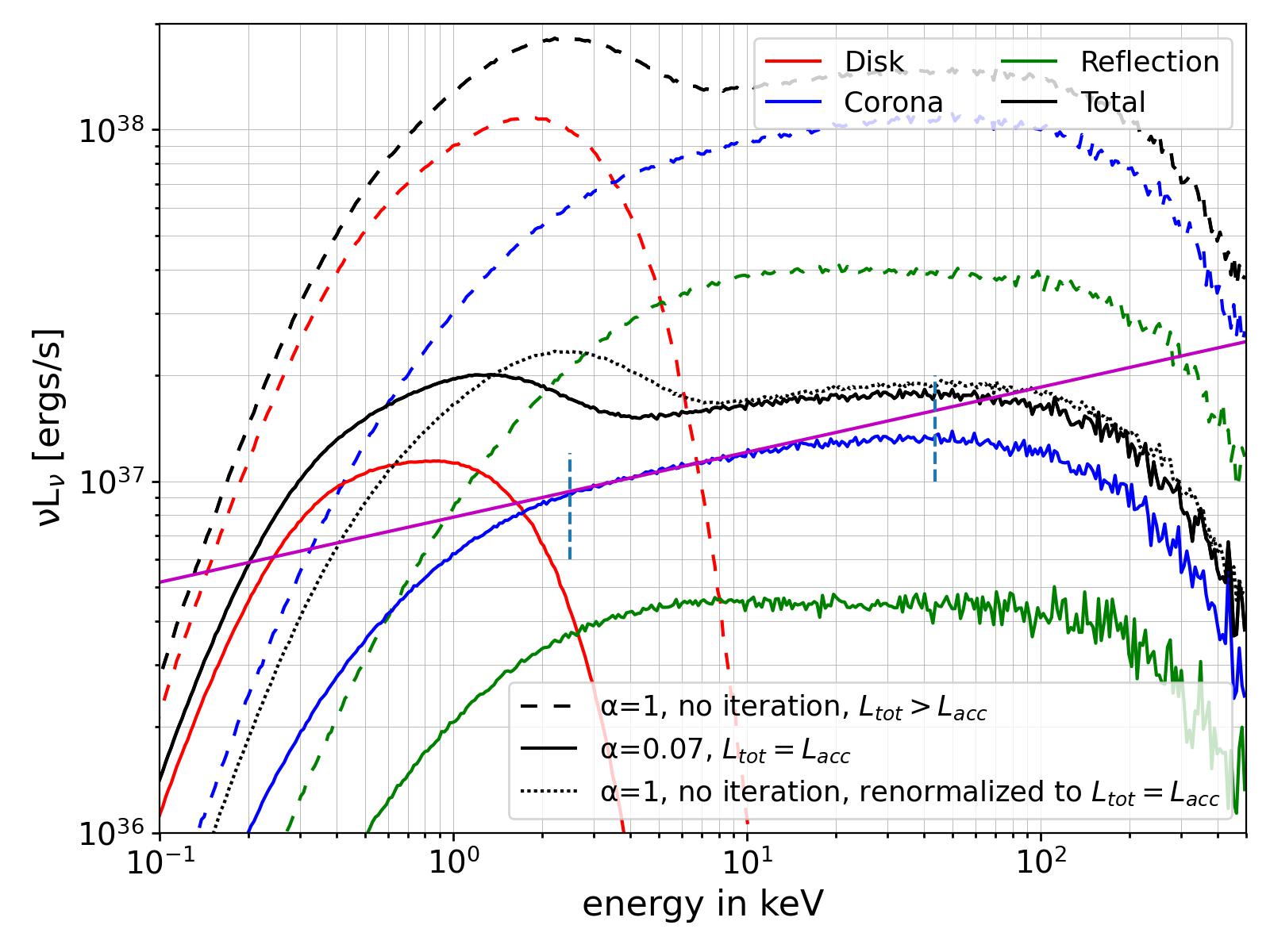}
\caption{Change in the observed spectrum at an inclination of 60$^\circ$ due to the imposition of global energy equilibrium ($L_{\rm tot} = L_{\rm acc}$). The coronal optical depth is fixed at 0.15, and all other parameters are set to their reference values as described in Section~\ref{sec_input_parameters}. Different spectral components are shown using distinct colors. The solid lines correspond to the equilibrium solution obtained through iteration of the disk dissipation fraction $\alpha$, which converges to a value of 0.07. The dashed lines represent the case with $\alpha = 1$, where no iteration is performed and the equilibrium condition $L_{\rm tot} = L_{\rm acc}$ is not enforced. The dotted black line shows the total spectrum for the $\alpha = 1$ case, but renormalized such that $L_{\rm tot} = L_{\rm acc}$. The solid magenta line represents a power law fit to the scattered spectrum of the equilibrium solution, yielding a photon index of 1.82. The two vertical lines mark the energy range of 2.5–43.7 keV over which the power law fit is performed.}
\label{fig_fraction1_consistent}
\end{figure}

Along with the reference set of parameters mentioned in Sect. \ref{sec_input_parameters}, here we also fix the optical depth of the corona $\tau=0.15$ and study the effect of imposing equilibrium on the disk-corona system. First we do not iterate $\alpha$ and fix its value to unity. For $\alpha=1$ case disk luminosity only due to intrinsic viscous dissipation becomes $L_{\rm acc}$ (equation \ref{eqn_L_acc}). In addition corona adds luminosity through inverse Comptonization and $L_{\rm tot}$ (equation \ref{eqn_total_luminosity_monk} after reaching convergence in luminosity as described in step (vii) of Sect. \ref{sec_procedure_steps}) becomes much larger than $L_{\rm acc}$. This is shown by the dashed lines in Fig. \ref{fig_fraction1_consistent}. Different colors represent different spectral components. To impose consistency of $L_{\rm tot}=L_{\rm acc}$ we iterate $\alpha$ (equation \ref{eqn_alpha_iteration}) and reach to the final equilibrium solution between disk and corona, spectral components of which are shown as the solid lines in Fig. \ref{fig_fraction1_consistent}. The final $\alpha$ for equilibrium decreases to 0.07 after iterations. The reduction in $\alpha$ to reach an equilibrium solution indicates that the luminosity contributed by the inverse Comptonization of seed photons by the hot coronal electrons is compensated by the reduction in luminosity radiated by the disk due to intrinsic dissipation. Basically (1-$\alpha$) fraction of $L_{\rm acc}$ is energizing the corona to keep it hot by balancing its cooling through inverse Comptonization. By dotted black line we also present the total spectrum for $\alpha=1$ case (no iteration on $\alpha$) but renormalized to $L_{\rm tot}=L_{\rm acc}$. The difference between the solid and dotted black lines show that the imposition of $L_{\rm tot}=L_{\rm acc}$ not only changes the luminosity but also the spectra. For the disk-corona in equilibrium with $\alpha=0.07$ ($\alpha=1$) the fraction of disk in the total luminosity at 60$^\circ$ inclination becomes 0.22 (0.24). If we consider the energy range 2-20 keV then the disk fraction becomes only 0.06 (0.23). 

\begin{figure*}
\centering\includegraphics[width=\textwidth]{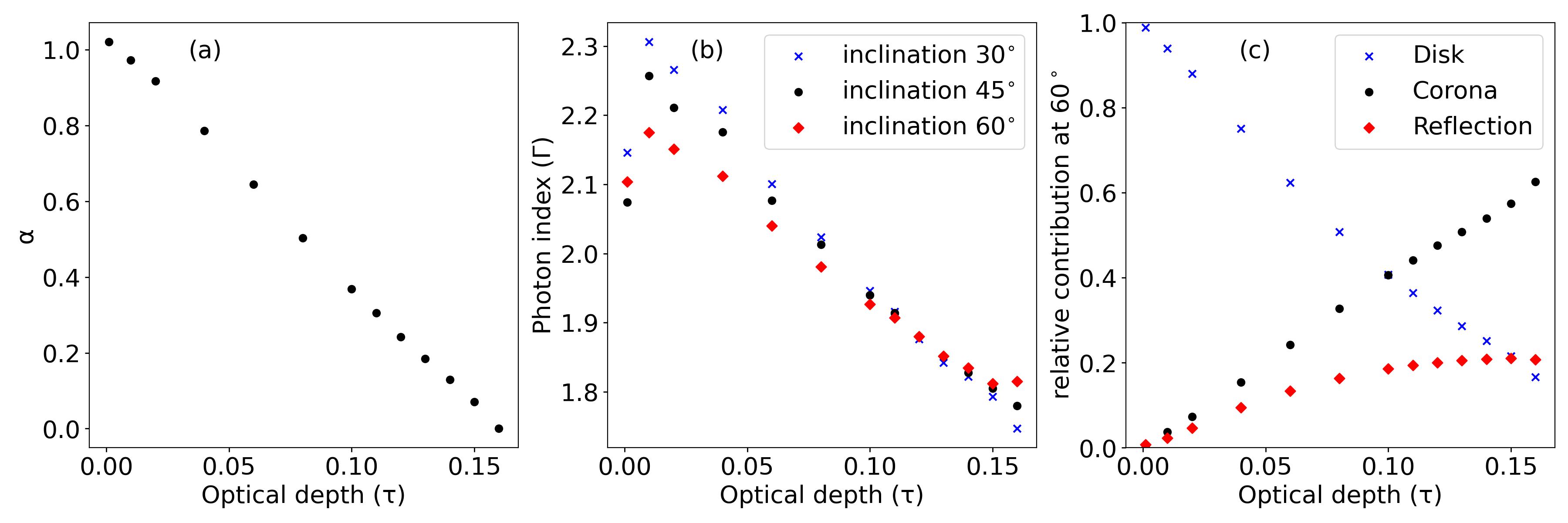}
\caption{Equilibrium solutions for increasing coronal optical depth, computed using the reference set of parameters. Panel (a): Variation of the disk dissipation fraction $\alpha$ with optical depth, illustrating a reduction in intrinsic disk dissipation as optical depth increases. Panel (b): Fitted photon index as a function of optical depth for different viewing inclinations. Panel (c): Fractional contributions of the disk, corona, and reflection components to the total luminosity (integrated over all energies) at an inclination of 60$^\circ$.}
\label{fig_alpha_flux_fraction}
\end{figure*}

We fit the scattered radiation with a power law following the method as described in Sect. \ref{sec_fitting_power_law} which is shown as the solid magenta line in Fig. \ref{fig_fraction1_consistent}. The energy range for fitting becomes 2.5-43.7 keV (4.4-58.1 keV for $\alpha=1$) which shifted toward lower energy for equilibrium solution because the seed blackbody from the disk itself shifted to lower energy due to low value of $\alpha$. The two vertical dashed lines indicate the energy range for the fitting. The fitted photon index at 60$^\circ$ inclination becomes 1.82. However, it is noticeable in Fig. \ref{fig_fraction1_consistent} that a single power law does not fit the higher-energy side very well although the energy range for fitting is computed appropriately where multiple scattering orders overlap with each other as described in Sect. \ref{sec_fitting_power_law}. We find this fact typical for other values of input parameters as well. The deviation of Comptonized radiation from a simple power law is well known as the anisotropy break when seed photons for Comptonization are coming from one side of the corona instead of injected in the mid-plane (Fig. 1 in \citealt{Petrucci2000}, Sect. \ref{sec_esc_aniso} in this draft). In addition to that, general relativistic effects may also contribute to the deviation.

\subsubsection{Limit on Comptonization}
\label{sec_bound_on_optical_depth}

To study the effect of optical depth -- and to determine the hardest spectra achievable for a given disk–corona system -- we fix all input parameters of the disk and corona to their reference values (as described in Section \ref{sec_input_parameters}), varying only the coronal optical depth. For each value of optical depth, we compute the corresponding equilibrium solution. Each point in Fig.~\ref{fig_alpha_flux_fraction} represents one such equilibrium, with the different panels illustrating how optical depth influences various output parameters. The accretion efficiency ($\eta$) and the total mass accretion rate onto the disk–corona system ($\dot{M}$) are held constant throughout, ensuring the total available accretion power ($L_{\rm acc}$) remains unchanged. As optical depth increases, the hot corona contributes more significantly to the total luminosity ($L_{\rm tot}$) due to enhanced Compton scattering. To maintain the energy balance required by the global equilibrium condition $L_{\rm tot} = L_{\rm acc}$ (see Section \ref{sec_procedure_steps}), this increase must be compensated by a decrease in other radiative components. This is achieved via a reduction in the disk dissipation fraction $\alpha$, found through our iterative scheme. A lower $\alpha$ leads to a reduced disk thermal radiation, thereby decreasing the seed blackbody photon input for Comptonization and counterbalancing the increased coronal output at higher optical depths. This behavior is shown in Fig.~\ref{fig_alpha_flux_fraction}(a), which demonstrates a gradual decrease in $\alpha$ with increasing $\tau$. However, if the corona becomes too optically thick, no physically meaningful value of $\alpha$ (i.e., $0\!\geq\!\alpha\!\geq\!1$) can restore equilibrium. In such cases, the system would require $\alpha < 0$ to satisfy $L_{\rm tot} = L_{\rm acc}$, which is unphysical. As a result, we find equilibrium solutions only for optical depths $\tau \lesssim 0.16$, corresponding to $\alpha \gtrsim 0$. This constraint imposes an upper limit on the allowed optical depth ($\tau_{\rm max}$), and consequently a lower limit on the photon index ($\Gamma_{\rm min}$), for any given disk–corona configuration.

The variation of the fitted photon index with increasing optical depth at different inclinations is shown in Fig.~\ref{fig_alpha_flux_fraction}(b). As optical depth increases, the coronal spectrum becomes progressively harder due to a greater number of scatterings. Additionally, the global equilibrium condition causes the seed blackbody spectrum for Comptonization to shift to lower energies, owing to the reduction in $\alpha$ with increasing optical depth. At very low optical depths, the number of scattered photons is small, leading to poor statistics and unreliable fitting. Moreover, with only one or two scatterings contributing significantly, the resulting spectrum exhibits a bumpy structure, making the fitted photon index for $\tau = 0.001$ an outlier. For an electron temperature of 120 keV and an albedo of 0.5, a static slab corona above the disk produces a Comptonized power law with photon index $\Gamma \gtrsim 1.7$. We find that at higher optical depths ($\tau > 0.11$), the Comptonized spectrum becomes harder at lower inclinations, consistent with previous studies (Table 1 of \citealt{Haardt1993}, Section 4 of \citealt{Zhang2019}). In contrast, for lower optical depths ($0.01 \leq \tau \leq 0.11$), the trend reverses: the spectrum becomes softer at lower inclinations. This change in behavior with inclination likely arises from the interplay between photons traveling vertically and those moving sideways through the corona. In our setup, the slab corona lies just above the disk, and each radial point of the disk emits blackbody photons isotropically in its local frame (equation 26 in \citealt{Zhang2019}). The specified optical depth refers to the vertical direction. Consequently, photons emitted at higher inclinations (i.e., at more oblique angles to the disk normal) traverse a longer path through the corona and thus experience a greater effective optical depth than those moving vertically. At low optical depths, sideways-traveling photons dominate the scattered radiation, leading to harder spectra at higher inclinations. As optical depth increases, vertically emitted photons begin to scatter more significantly, reversing the trend. This shift in the dominant direction of seed photon scattering is responsible for the inclination-dependent transition in the spectral hardness behavior.

For a given disk–corona configuration, the global equilibrium condition not only determines the specific values of $\alpha$ and $\Gamma$, but also uniquely sets the fractional luminosity contributions from the disk, corona, and reflection components. Figure~\ref{fig_alpha_flux_fraction}(c) illustrates how these relative contributions vary with optical depth at an inclination of 60$^\circ$, integrated over the full energy band. As the optical depth increases, the disk’s contribution to the total luminosity decreases, compensating for the growing emission from the corona. Interestingly, while the coronal luminosity continues to rise with increasing optical depth, the reflection component shows an early steep increase followed by saturation. This behavior arises because at high optical depths a significant fraction of the reflected photons are unable to escape the corona and instead undergo additional scatterings along with the disk blackbody photons. In this computation, we adopt an albedo of 0.5, which leads to nearly equal contributions from the disk and reflection components at $\tau = \tau_{\rm max} = 0.16$. At this point, $\alpha \sim 0$, implying that intrinsic dissipation within the disk is nearly negligible. As a result, all the disk emission originates from the reprocessing of coronal illumination rather than from internal viscous heating.

\subsubsection{Effect of BH spin, coronal temperature and albedo}
\label{sec_effect_spin_Te_albedo}

\begin{figure*}
\centering\includegraphics[width=\textwidth]{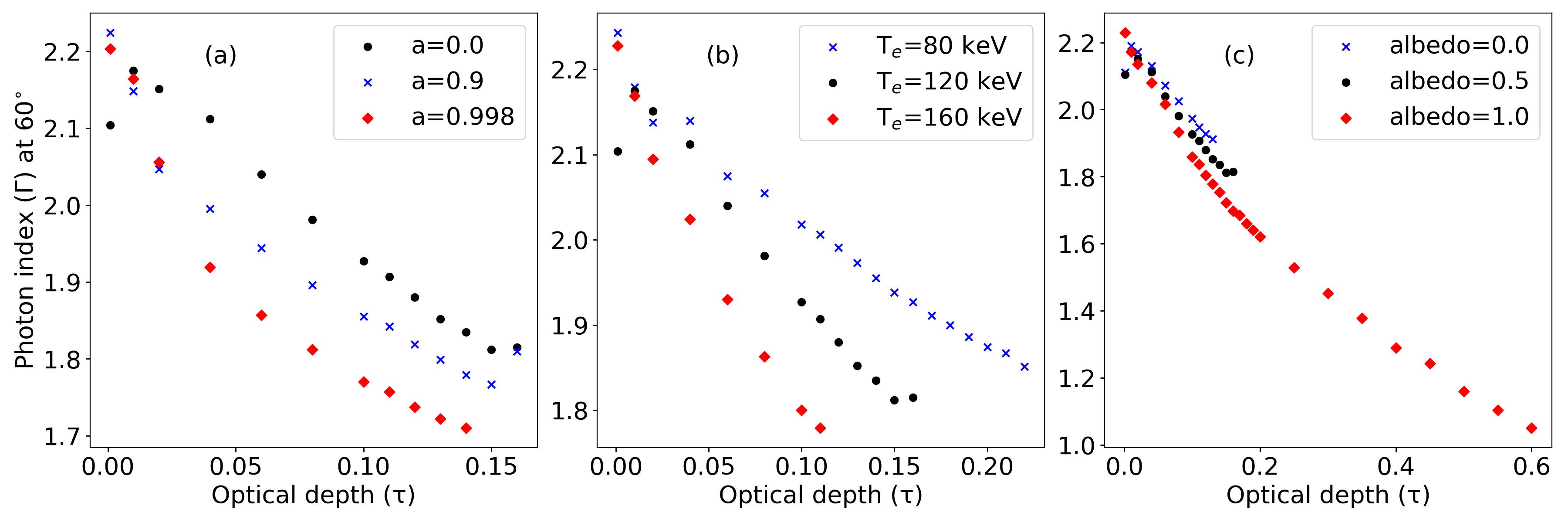}
\caption{Effect of different parameters on the spectral hardness of equilibrium solutions at an inclination of 60$^\circ$. In each panel, a single parameter is varied while keeping all others fixed at their reference values, as described in Section~\ref{sec_input_parameters}. The reference parameter set (black hole spin $a = 0$, coronal electron temperature $T_{\rm e} = 120$ keV, and disk albedo = 0.5) is shown by black circles in all panels. Panel (a), (b), and (c) shows the effect of black hole spin, coronal electron temperature, and disk albedo respectively. In general, higher spin, higher electron temperature, and higher albedo lead to harder spectra at a given optical depth. However, among these, only increasing the albedo results in a significant reduction of the minimum photon index $\Gamma_{\rm min}$.}
\label{fig_effect_of_spin_Te_albedo}
\end{figure*}

In addition to the ${\rm a}=0$ case presented before as reference set in Sect. \ref{sec_bound_on_optical_depth}, we find the equilibrium solution with ${\rm a}=0.9$ and 0.998 keeping all the other parameters same as in Sect. \ref{sec_input_parameters}. For different spins, accretion efficiency becomes different which makes the mass accretion rate $\dot{M}=0.1\dot{M}_{Edd}$ different in physical units but keeps the available accretion power same ($L_{\rm acc}=6.2\times10^{37}$ erg/s). Fig. \ref{fig_effect_of_spin_Te_albedo}(a) represents the variation of photon index at $60^\circ$ inclination with the BH spin. At the same optical depth of the corona, we obtain harder spectra with higher spin ($\Delta\Gamma\sim-0.15$ for $\Delta {\rm a}=0.998$). Although for all the spins total accretion power remains the same, due to the larger area of the disk (as the inner boundary of the NT disk, ISCO, moves closer to the BH for higher spin) and the increased light bending effect, the disk illumination eventually becomes larger. Larger illumination increases the reprocessed component of the blackbody entering to the corona. This results to larger coronal output as well. For higher spin, the luminosity lost in the BH from both disk and corona increases. This increment in different luminosity components (equation \ref{eqn_total_luminosity_monk}) for higher spin forces $\alpha$ to be lower (equation \ref{eqn_alpha_iteration}) to bring the disk-corona in equilibrium ($L_{\rm tot}=L_{\rm acc}$). This reduction in $\alpha$ decreases the contribution from seed blackbody which makes the spectra harder with higher spin. The difference in photon indexes for different spins is larger at higher inclinations (not shown in the plots). If we increase the coronal temperature, then the spectra also become harder. Fig. \ref{fig_effect_of_spin_Te_albedo}(b) shows the variation of photon index at 60$^\circ$ inclination with electron temperature. Larger electron temperature transfers larger energy through inverse Comptonization, making the coronal output higher and $\alpha$ lower for equilibrium solution, subsequently the spectra harder. However, although higher spin and electron temperature produces relatively harder spectra at the same optical depth, only little improvement is achieved in lowering the $\Gamma_{\rm min}$. The trend of hardening the spectra with higher BH spin remains the same even when the coronal temperature is estimated from GRMHD simulation, resulting larger temperature for higher spin \citep{Kinch2021spin}.

Figure~\ref{fig_effect_of_spin_Te_albedo}(c) shows how the photon index ($\Gamma$) of the Comptonized spectrum at 60$^\circ$ inclination varies with albedo. In addition to the previously discussed case of albedo = 0.5, we compute spectra for albedo = 0 (complete absorption) and albedo = 1 (complete reflection), keeping all other parameters fixed to their reference values (section~\ref{sec_input_parameters}). Consistent with earlier studies (e.g., Fig.~4 in \citealt{Malzac2005}, Fig.~3 in \citealt{Poutanen2018}), we find that higher albedo leads to systematically harder spectra. For albedo = 1, we halt the increase in optical depth at $\tau = 0.6$, where the resulting spectrum reaches $\Gamma \sim 1.0$. The dependence of spectral hardness on albedo arises primarily from its effect on the reprocessed blackbody radiation from the disk, rather than from intrinsic dissipation. A higher albedo reduces the amount of absorbed and reprocessed flux, thereby lowering the soft seed photon population for Comptonization. This leads to harder spectra at a fixed optical depth and significantly lowers the minimum achievable photon index ($\Gamma_{\rm min}$). In the case of high optical depth ($\tau > \tau_{\rm max}$), the increase in reprocessed blackbody flux—driven by stronger coronal illumination—can no longer be offset by reducing $\alpha$, as it has already reached zero. This prevents the system from attaining equilibrium. However, when albedo is increased, the reprocessed blackbody component is suppressed, allowing equilibrium to be reached even at higher $\tau$ values. For albedo = 1, the reprocessed blackbody contribution vanishes entirely, effectively decoupling the corona from the disk. In this scenario, the seed blackbody photons arise solely from intrinsic dissipation (equation~\ref{eqn_Teff}). At large optical depths, the corresponding low $\alpha$ values yield minimal seed photon flux, resulting in only a modest increase in coronal luminosity with increasing $\tau$. This reduced feedback allows a much larger $\tau_{\rm max}$ and, consequently, a much lower $\Gamma_{\rm min}$.

\subsection{Escaping and anisotropic fraction}
\label{sec_esc_aniso}

\begin{figure}
\centering\includegraphics[width=\columnwidth]{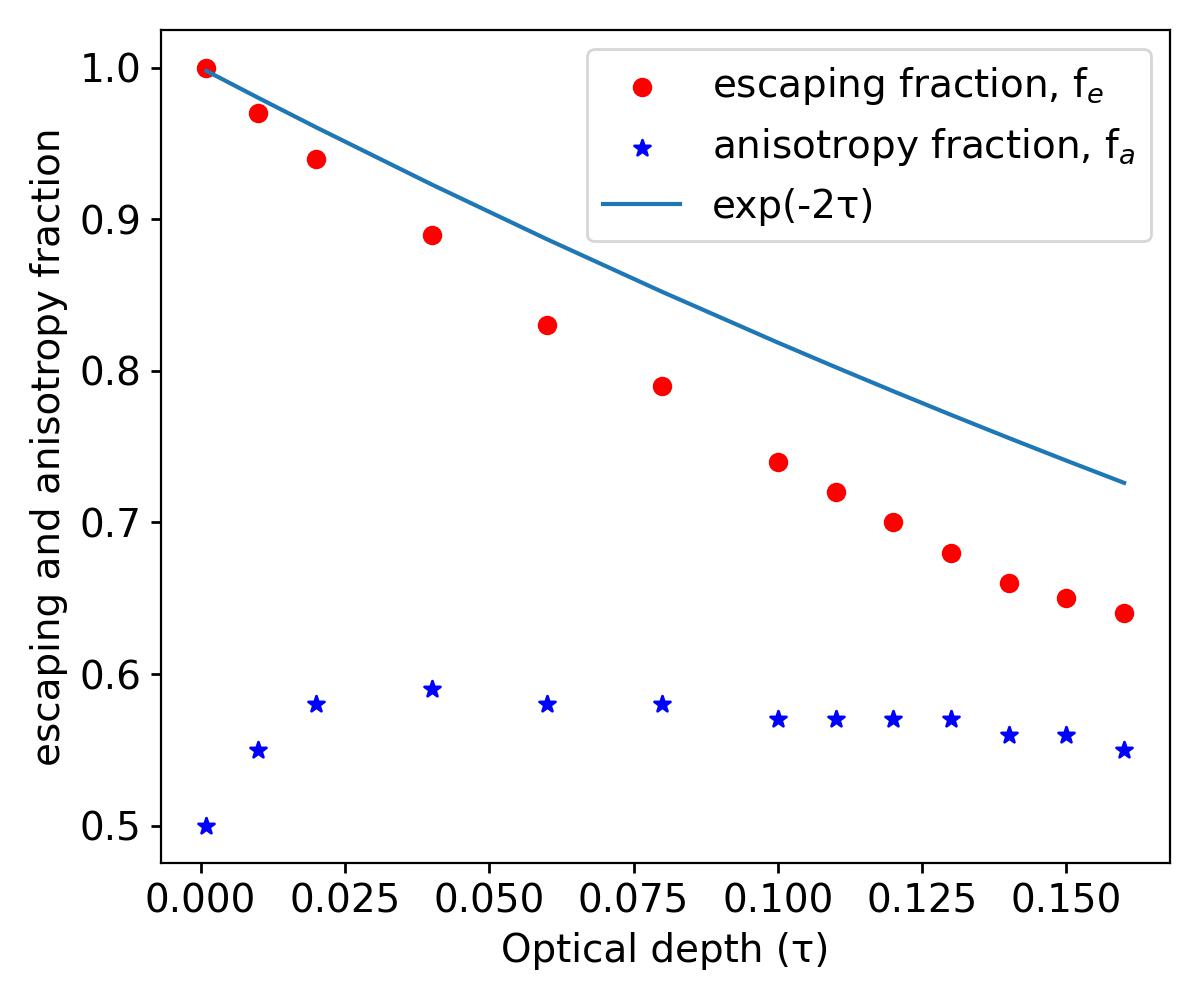}
\caption{Variation of the escaping fraction ($f_{\rm e}$) and anisotropic fraction ($f_{\rm a}$) with the coronal optical depth ($\tau$). The equilibrium solutions are computed using the same reference parameter set as in Fig.~\ref{fig_alpha_flux_fraction}. To facilitate comparison, the theoretical expectation for $f_{\rm e}$, given by $\exp(-2\tau)$, is also shown. The factor of 2 arises because $\tau$ in MONK is defined as the vertical optical depth from the mid-plane of the slab to its surface.}
\label{fig_leak_aniso}
\end{figure}

The equilibrium solution for a given disk–corona system is achieved through iteration of the $\alpha$ parameter, which represents the fraction of the total available accretion power dissipated in the disk. Once equilibrium is established, the final converged value of $\alpha$ not only fixes the intrinsic disk and coronal luminosities, but also determines the distribution of all radiative components. In particular, it allows us to compute two important diagnostic quantities: the escaping fraction ($f_{\rm e}$) and the anisotropic fraction ($f_{\rm a}$). The escaping fraction $f_{\rm e}$ quantifies the fraction of the total disk luminosity that escapes directly to infinity without undergoing any scattering. The anisotropic fraction $f_{\rm a}$ is defined as the fraction of scattered coronal luminosity that irradiates the disk. The mathematical definitions of both quantities are provided in Appendix~\ref{appendix_theoretical_estimates}. Since the MONK computation tracks each individual photon contributing to the global disk–corona equilibrium, we are able to determine the luminosity in each component with high accuracy. From these, we compute $f_{\rm e}$ and $f_{\rm a}$ by taking appropriate ratios of the relevant luminosities after the equilibrium solution has been obtained.

For a slab corona with vertical optical depth $\tau$, it is instructive to compare the actual escaping fraction $f_{\rm e}$ with the naive expectation $\exp(-\tau)$. We find that $f_{\rm e}$ is systematically lower than $\exp(-\tau)$ due to geometric and angular effects. Specifically, photons do not all propagate vertically; those emitted at oblique angles traverse a larger effective optical depth and are thus more likely to be scattered. Moreover, the angular distribution of the emitted photons is not isotropic—limb darkening and the $\cos\theta$ dependence of the emission introduce directional weighting, further reducing $f_{\rm e}$ relative to the vertical approximation. The anisotropic fraction $f_{\rm a}$ also reflects the inherent geometry of the system. Since the seed blackbody photons are emitted from the bottom boundary of the slab corona (i.e., the disk surface) rather than from its mid-plane, the setup naturally introduces an asymmetry in the radiation field. Photons that are scattered backward toward the disk typically undergo more energetic scatterings compared to those scattered forward toward infinity. As a result, a larger fraction of the scattered luminosity tends to illuminate the disk rather than escape, as also highlighted in \citet{Haardt1993anisotropy}. This anisotropy plays a crucial role in shaping the balance of energy between the disk and the corona.

For the equilibrium solutions obtained using the reference parameter set described in Section~\ref{sec_input_parameters}, we present the variation of the escaping fraction ($f_{\rm e}$) and anisotropic fraction ($f_{\rm a}$) as functions of optical depth $\tau$ in Fig.~\ref{fig_leak_aniso}. To assess the deviation of $f_{\rm e}$ from analytical expectations, we also plot $\exp(-2\tau)$. We note that the factor of 2 arises because in MONK, $\tau$ is defined as the vertical optical depth from the mid-plane to the boundary of the slab. As anticipated, $f_{\rm e}$ falls below this analytical estimate due to angular dependence and geometric effects, as discussed earlier. The behavior of $f_{\rm a}$ with $\tau$ is more complex. At low optical depths ($\tau \leq 0.025$), the illumination of the disk is dominated by singly scattered photons, which tend to be backward-scattered and hence strongly anisotropic. As $\tau$ increases in this regime, more photons are scattered once and redirected toward the disk, leading to an increase in $f_{\rm a}$. However, at higher optical depths ($\tau > 0.025$), multiple scatterings become significant. These multiple scattered photons are more isotropic in their angular distribution, which slightly reduces $f_{\rm a}$. This trend reflects the reduction of anisotropy with increasing scatterings, a phenomenon that also manifests as the so-called `anisotropy break' in the Comptonized spectrum \citep{Petrucci2000}. Although we only show results up to $\tau = 0.16$—beyond which no equilibrium solution with $L_{\rm tot} = L_{\rm acc}$ is possible for our setup—the qualitative behavior of $f_{\rm a}$ continues to evolve at larger optical depths. For $\tau \gtrsim 5$ (with the exact threshold depending on the electron temperature), $f_{\rm a}$ begins to increase monotonically, eventually approaching unity, as expected for a slab that fully traps and reprocesses the radiation.

\subsection{Indication from local imbalance}
\label{sec_local_global_balance}

\begin{figure}
\centering\includegraphics[width=\columnwidth]{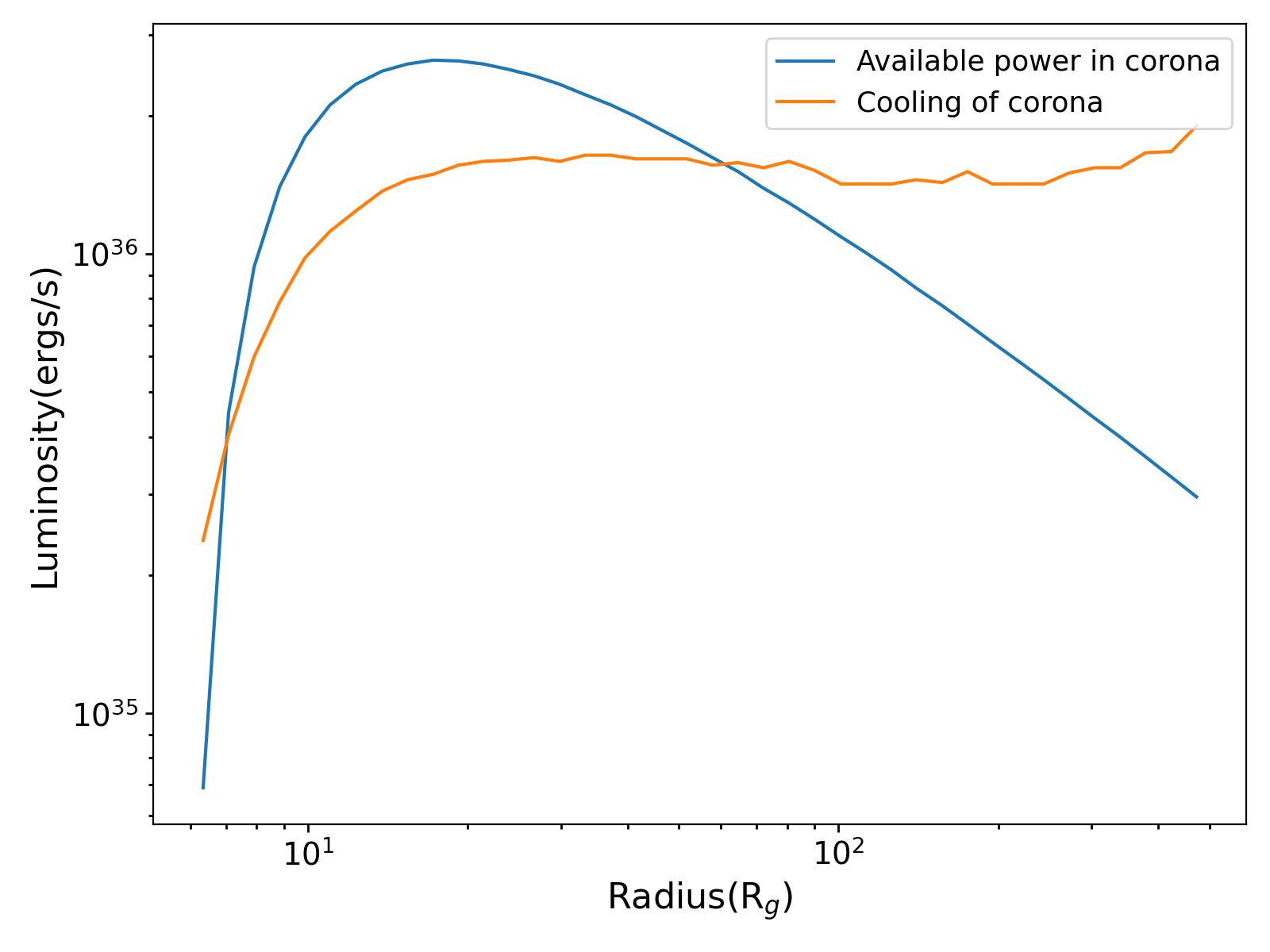}
\caption{Comparison of the coronal cooling rate with the available accretion power in the corona as a function of radius, for a coronal optical depth of $\tau = 0.15$. All other parameters are set to their reference values.}
\label{fig_cooling_corona_available}
\end{figure}

In our computations, we consider an extended slab corona with uniform physical properties—specifically, constant electron temperature and optical depth across all radial locations—positioned just above the accretion disk. With this configuration, we impose a condition of global energy equilibrium, requiring that the total luminosity emitted by the disk-corona system matches the total available accretion power. However, we do not impose any condition of local equilibrium, i.e., energy balance between the disk and corona at each radius. To examine the consistency of the global equilibrium solution, we compare the radial profile of coronal cooling with the available local heating power in the corona. The radial distribution of coronal cooling is computed by tracking the energy transferred from electrons to photons during each scattering event within the corona, recording both the scattering location and the energy exchange. This information is then vertically integrated over each radial annulus—defined identically to the disk annuli—to obtain the total cooling rate per radial bin. The available heating power for the corona at each radius is estimated as $(1-\alpha)$ times the local accretion energy release given by the Novikov-Thorne (NT) prescription. That is, the fraction $(1-\alpha)$ of the NT flux is assumed to heat the corona locally, while the remaining $\alpha$ contributes to the intrinsic disk dissipation. As per the NT model, the luminosity drops sharply at the innermost stable circular orbit (ISCO) due to the zero-torque boundary condition \citep{Novikov1973, Zimmerman2005}, and this feature is naturally incorporated in our computation of local accretion power.

The radial variation of the power available in the corona and the corresponding coronal cooling rate is shown in Fig.\ref{fig_cooling_corona_available} for $\tau = 0.15$, with all other parameters fixed to the reference set (Section\ref{sec_input_parameters}). Global energy balance ensures that the total area under both curves equals the available coronal power, i.e., $(1 - \alpha) \times L_{\rm acc} = 5.6 \times 10^{37}$ erg s$^{-1}$, where the equilibrium solution yields $\alpha = 0.07$ (see Section~\ref{sec_effect_of_equilibrium}, Fig.~\ref{fig_fraction1_consistent}). We observe that in most of the inner region of the slab corona, the cooling rate lies below the locally available power, implying that a portion of the dissipated energy remains stored in the corona. In contrast, within a narrow innermost zone and across the entire outer region, the coronal cooling rate exceeds the locally available power. While the discrepancy in the innermost region is relatively small—and sensitive to the radial binning scheme—the excess cooling in the outer regions is substantial. This mismatch can be primarily attributed to our simplifying assumption of a uniform electron temperature and optical depth across the slab. Physically, we expect the electron temperature to decrease with increasing radius, as the corona becomes less compact and less energetic away from the black hole. Such a radial decline in temperature would naturally reduce the cooling rate in the outer regions, potentially bringing it in line with the locally available power. Therefore, to achieve not only global but also local energy balance between the disk and the corona, either (i) the electron temperature must decrease with radius, or (ii) there must be a mechanism—possibly magnetic energy transport—that redistributes energy from the inner to the outer corona to sustain the high temperatures required there. In future work, we will explore radial profiles of decreasing electron temperature to examine whether such configurations can produce self-consistent solutions satisfying both global and local energy balance. 

\section{Discussion}
\label{sec_discussion}
\subsection{Static slab corona (possibly) can not explain hard state observations}
\label{sec_slab_corona_hard_state}

Assuming a disk albedo of 0.5 and a typical coronal electron temperature, we find that the hardest photon index ($\Gamma_{\rm min}$) achievable with a static slab corona fully covering the disk is approximately 1.7–1.8, depending on the black hole spin, coronal temperature, and inclination. Although the relationship between disk albedo and ionization parameter ($\xi$) is not straightforward, Fig. 3 of \citet{Malzac2005}—which estimates albedo based on reflected/reprocessed flux above 1 keV—suggests that an albedo of 0.5 corresponds to $\xi \sim 2000$ for a constant-density slab. Thus, the $\Gamma_{\rm min}$ values obtained in our study would be expected only when $\xi \sim 2000$, which is consistent with earlier studies of two-phase accretion flows in local energy balance \citep{Haardt1993, Malzac2005, Poutanen2018}. Observationally, typical ionization parameters for the inner disk reflection fall in the range $\xi \sim 10^3$–$10^4$, supporting the plausibility of our assumed albedo \citep{Basak2017, Steiner2017, Buisson2019, Chakraborty2020, Zdziarski2021MAXI, Chand2024}. Even much more self-consistent treatment of disk-corona equilibrium incorporating GRMHD simulation, non-uniform temperature and optical depth of the corona, and energy dependent disk albedo with proper photoionization treatment can produce the hardest power law only $\sim$ 1.6 \citep{LiuKrolik2025}. However, for most X-ray binaries, the hardest observed photon indices lie in the range $\Gamma \sim 1.5$–2.0 \citep{Wu2008, Yang2015, Liu2023, You2023}, with some requiring $\Gamma < 1.5$ \citep{Chakraborty2020, Datta2024, Chand2024}. Therefore, while spectra with $\Gamma \sim 1.6-1.7$ can be reproduced with a static slab corona above the disk, harder spectra likely require an alternative mechanism or coronal geometry. One possible way to produce harder spectra is by increasing the disk albedo (and hence its ionization), as shown in Fig.~\ref{fig_effect_of_spin_Te_albedo}(c). However, this comes at the cost of significantly reducing or even washing out the iron line feature in the reflection spectrum. This might be acceptable if the observed iron line in the hard state is primarily produced by outer disk reflection, while the inner disk reflection contributes mainly to the continuum. Such an interpretation is supported by the top panel of Fig.~5 in \citet{Zdziarski2021MAXI} and Fig.~3 in \citet{Datta2024}. This scenario is also consistent with the observed weakening of the iron line in the hard state, possibly due to reflection passing through a Comptonizing corona before reaching the observer \citep{Petrucci2001, Steiner2016}. Conversely, if a single reflection component suffices, or if both inner and outer reflection components are required to reproduce the iron line feature—as suggested by \citet{Liu2023}—then there is no flexibility to arbitrarily increase the disk ionization or albedo. In such cases, the static slab corona cannot explain spectra with $\Gamma < 1.7$, effectively ruling it out as a viable geometry for the hardest observed spectral states in XRBs. 

The key requirement for producing harder spectra lies in suppressing the soft blackbody component. This can be achieved through either (i) a truncated disk geometry with little or no overlap between the disk and corona \citep{Poutanen2018}, and/or (ii) an outflowing corona \citep{Malzac2001}. In the case of a truncated disk, nonthermal seed photons—such as those from bremsstrahlung or synchrotron emission—become essential for Comptonization if the disk and corona do not overlap; otherwise, the resulting spectrum can become excessively hard. With the framework developed in the current version of MONK, we plan to investigate these alternative geometries in a follow-up study. Beyond spectral hardness, the fractional contribution of different emission components to the final spectrum (e.g., Fig.~\ref{fig_alpha_flux_fraction}(c)) provides additional constraints on the disk–corona geometry and associated model parameters.

\subsection{Limitations to include atomic transitions for reflection}
\label{sec_limitation_refl_table}
One of the major limitations of our model is that the current reflection treatment does not include atomic transitions, which prevents us from fitting observed reflection spectra accurately. Although it is in principle possible to couple our Monte Carlo code with an external radiative transfer code to model reflection more realistically—an approach demonstrated by \citet{Malzac2005, Kinch2019ptransx}—this would significantly increase computational complexity and resource requirements, and is therefore beyond the scope of the present work. An alternative approach would be to adopt reflection spectra from pre-computed reflection tables, such as {\tt reflionx} \citep{Ross2005}, {\tt XILLVER} \citep{Garcia2010, Garcia2013}, or the more recent STOKES tables that include polarization \citep{Podgorny2022}. This methodology has been employed in previous works, including \citet{Poutanen2018} and \citet{Krawczynski2022kerrC}. However, there are several reasons that currently prevent us from incorporating these tables into our framework.

First and foremost, among currently available models, only a single reflection table—{\tt refhiden} \citep{Ross2007}—includes both intrinsic viscous dissipation and external power law illumination. Even this model is limited: it assumes a minimum blackbody temperature of 0.4 keV, making it suitable only for very high mass accretion rates onto stellar-mass black holes. None of the other widely used reflection tables incorporate intrinsic disk dissipation, which has been shown to significantly affect the reflection features (see Figure 6 in \citealt{Ross2007}). Additionally, these tables do not account for self-irradiation of the disk, a factor that can become important for high-spin black holes and low coronal heights \citep{Mirzaev2024}. Furthermore, when modeling reflection that includes many atomic transitions alongside scattering processes, it becomes computationally challenging to achieve a fully converged solution for the temperature and radiation field within a reasonable number of iterations. This often results in incomplete convergence and a mismatch between the incident and reflected flux, limiting the reliability of these tables to only a narrow range of input parameters (see Section 2.1 of \citealt{Poutanen2018}; footnote 1 in \citealt{Zdziarski2020}; and Section 3.3 in \citealt{Krawczynski2022kerrC}). A final and fundamental limitation arises from the specific assumptions about illumination in these reflection tables. They typically assume a power law spectrum with fixed high- and low-energy cutoffs as the incident radiation. To use such tables in our framework, we would need to approximate the incident radiation at each radial annulus of the disk as a power law with a cutoff. This poses a significant computational challenge: for an optically thin corona, the number of photons diminishes rapidly with the number of scatterings, making it particularly difficult to characterize the high-energy cutoff accurately. Since reflection depends not only on photon energy but also on incidence and emission angles, we would need to model angular distributions of the incoming radiation, dramatically increasing the required photon statistics. Therefore, while incorporating reflection spectra with atomic transitions remains an important future direction, it demands substantial methodological and computational developments beyond the scope of the present study.

\subsection{Effect of self-irradiation without corona}
\label{sec_illumination_selfirr}

If we consider only the disk and its self-irradiation and no corona, then we expect that for equilibrium solution the total luminosity collected at infinity ($L_{\rm tot}$ in equation~\ref{eqn_total_luminosity_monk}) to be $L_{\rm acc}$, i.e., $\alpha$ should be equal to 1 as there is no additional luminosity contributed by the corona. However, it does not become strictly true always. As the self-irradiation becomes more significant with higher spin, $\alpha$ deviates more from 1. The reasons for this discrepancy are two fold. First one is the ignorance of second order reflection (reflection of reflected photons) in our computation which becomes more relevant at higher albedo. Second one is the ignorance of work done by the self-irradiation on the disk. We keep the angular momentum profile of the disk as prescribed by \cite{Novikov1973} in Kerr geometry even with self-irradiation. Due to this imposition, there can be slight gain or loss of energy of a photon due to reflection with respect to an observer at infinity. Because the gravitational red-shift of a photon depends not only on the location of the disk but also on the direction of the photon propagation. Therefore, due to diffuse nature of reflection and reprocessed radiation, the average direction of photon after reflection or absorption of self-irradiation may change. This effect may lead to a change in energy flux with respect to an observer at infinity even though Thomson scattering through Chandrasekhar's semi-infinite electron atmosphere conserves energy flux in the comoving frame of the disk. To model the effect of self-irradiation precisely, we need to solve the hydrodynamics of the Keplerian disk incorporating the angular momentum transport by the self-irradiation as incorporated in KERRBB model \citep{Li2005}. However, this effect is always quite small with respect to the flux irradiated to the disk at every radii (Fig. 1 in \citealt{Li2005}). With no corona, due to self-irradiation we find that $\alpha$ can differ from 1 by 5\% in the worst case scenario. The effect of self-irradiation is relevant only when either there is no corona or it is very optically thin. For our main focus of this work, with disk-corona interaction, we can safely ignore this effect as coronal dissipation plays a  major role in achieving the equilibrium.

\subsection{Comparison with methodology of HM93 (\citealt{Haardt1993})}
For a two-phase accretion flow (disk and corona) in equilibrium, we present in Appendix \ref{appendix_theoretical_estimates} a set of analytical estimates for various luminosity components in the Newtonian framework, following the approach of HM93. However, there is a fundamental difference between our formalism and HM93. We introduce the escaping fraction parameter $f_{\rm e}$, and assume that the coronal luminosity is the sum of the disk luminosity which interacts with the hot electrons and the fraction of accretion power available to energize the hot electrons (equation \ref{eqn_L_corona}). On the other hand, HM93 assumes that the coronal luminosity is only the accretion power available in the corona. These two assumptions converge with the approximation of $f_{\rm e}=1$, leading to the same analytical estimations of luminosities in different radiative components through our formalism and HM93 as shown at the end of Appendix \ref{appendix_theoretical_estimates}. From Fig. 1(b) of HM93 it is clear that the hardest photon index achievable for two phase accretion flow in equilibrium is for the optically thinner corona which has higher electron temperature. For very optically thin corona, the escaping fraction of luminosity $f_{\rm e}$ will become very close to unity and nullify any difference between our presented analytical estimations and HM93.

In our numerical framework, equilibrium solutions are obtained by iterating the parameter $\alpha$ within the MONK code under the constraint $L_{\rm tot} = L_{\rm acc}$. From these simulations, we obtain the fractions of various luminosity components observed at infinity (e.g., as shown in Fig. \ref{fig_alpha_flux_fraction}(c) for an inclination of 60$^\circ$). We then use the analytic expressions from Appendix \ref{appendix_theoretical_estimates}, inputting the corresponding values of $\alpha$, $f_{\rm e}$, $f_{\rm a}$, and the assumed albedo, to independently compute the fractional luminosities. These analytic predictions agree very well with the values obtained directly from MONK, particularly for low black hole spin, where general relativistic corrections are minimal and thus not required in the Newtonian estimates.

It is worth noting that the computation of $f_{\rm e}$ and $f_{\rm a}$ from the MONK output involves only the luminosities entering and escaping the corona, as well as the scattered luminosity from the corona illuminating the disk. The condition used in the iterative scheme, $L_{\rm tot} = L_{\rm acc}$, does not explicitly involve the disk emission luminosity (e.g., as defined in equation \ref{eqn_disk_emission}). Therefore, the good agreement between the full set of luminosity components derived analytically and those obtained from MONK validates the self-consistent iteration scheme used in our approach. The primary purpose of the analytic estimates is to demonstrate that, although our numerical method reaches the equilibrium solution by iterating on the total luminosity $L_{\rm tot}$ alone (step (ix) in Section 2), the resulting solution naturally aligns with the analytical predictions for individual luminosity components, similar to the framework in HM93. In our case, inverse Comptonization and the additional power provided by the hot corona are handled entirely by MONK, and the equilibrium is found without requiring the analytic luminosity expressions. In contrast, HM93 modeled Comptonization using an amplification factor and relied on analytic luminosity estimates to relate this factor to the spectral slope and the fraction of energy dissipated in the corona.

\subsection{Similarities and differences with other models}
\label{sec_kynsed_kerrc}
\citet{Dovciak2022} developed a physical model, {\tt kynsed}, to study the global energy balance between the disk and the corona, enabling broadband SED fitting of AGNs within {\tt xspec}. In this model, self-consistency is achieved by transferring a fraction of the accretion energy to a point-like, lamp-post corona, in accordance with the observed power law component. The disk is then heated via illumination by this corona. In contrast to our approach—where a uniform fraction of accretion energy is transferred from the disk to the corona at all radii—{\tt kynsed} assumes that only the inner region of the disk, within a specified transition radius, transfers its entire energy to the corona. Reflection is modeled using the {\tt xillver} tables \citep{Garcia2013}, and the absorbed flux is estimated by subtracting the reflected flux (above 0.1 keV) from the incident illumination. Another key difference lies in the treatment of Comptonization. In {\tt kynsed}, the corona is assumed to be a point-like lamp-post located along the disk axis at a certain height, and its emission is modeled as a power law with a given normalization—no actual Comptonization is computed. After fitting, the model estimates the size and optical depth of the corona by assuming a spherical geometry. In contrast, MONK performs full Comptonization from an extended corona and, as shown in \citet{Zhang2024}, this significantly alters the disk illumination profile compared to that of a lamp-post geometry. However, MONK does not currently support direct spectral fitting of observational data unless a spectral table is generated from its output.

Another contemporary model, {\tt kerrC}, developed by \citet{Krawczynski2022kerrC}, performs ray tracing in the Kerr metric and models Comptonization by an extended corona—conceptually similar to MONK. However, while it includes reflection from the disk, it does not perform any iterative feedback between the disk and corona and thus does not achieve a final equilibrium solution. A key distinction lies in how reflection is modeled. {\tt kerrC} implements reflection using Chandrasekhar’s multiple scattering formalism, as done in our work. However, it then rescales the weight of each photon based on the {\tt xillver} high-density reflection tables \citep{Garcia2016}, under the assumption that the reflection spectrum from a highly ionized slab matches that of a pure electron atmosphere. This approximation enables {\tt kerrC} to estimate the polarization of reflected emission using the {\tt xillver} tables, which themselves do not include polarization information. However, recent studies on reflection with polarization suggest that emission from atomic transitions leads to significant depolarization, even in highly ionized slabs \citep{Podgorny2022}. This implies that the scaling approach adopted by {\tt kerrC} may not be appropriate for accurately modeling polarized reflection.

It should be noted that with the current upgrade of the MONK code, it becomes capable to find equilibrium solution for a given disk-corona system incorporating exact prescription of Comptonization by extended corona in complete relativistic framework. This is novel in nature and no other available code does the similar computation as far as we know. Therefore, although the previous version of MONK code \citep{Zhang2019} is benchmarked with {\tt kynsed} (Fig. 6 in \citealt{Dovciak2022}) and {\tt kerrC} (section 3.5 in \citealt{Krawczynski2022kerrC}), it is not possible to compare the results from these codes with the equilibrium solutions achieved through the upgraded version of MONK.

\subsection{Possibilities to improve}

In this first attempt to find an equilibrium solution with the disk in global balance with an extended corona, several simplifying assumptions have been made. Most notably, we assume a homogeneous corona with uniform optical depth and electron temperature across all radii. However, in reality, the electron temperature is expected to decrease with increasing distance from the black hole. While the density should also decrease outward, the accompanying increase in scale height makes it nontrivial to predict how the coronal optical depth varies with radius. Another simplification is the heating of the disk in addition to the Novikov–Thorne (NT) profile, without solving the disk’s hydrodynamics. This is clearly an idealization. A more realistic treatment would involve using radial and vertical profiles of temperature and optical depth derived from analytical models or simulations of two-phase accretion flows \citep{Cho2022, Bambic2024}, though such models are still limited in availability. We also assume, somewhat ad hoc, that the same fraction of accretion energy is available to the corona at all radii, which may not reflect the actual physical scenario. In Section \ref{sec_local_global_balance}, we discuss how enforcing local equilibrium between the corona and the underlying disk at each radius may offer insights into the radial variation of $\alpha$ or the inhomogeneous nature of the corona. Additionally, for an extended corona above the disk, a typical emissivity index of 3 implies that the disk ionization should decrease with radius. This would, in turn, reduce the albedo at larger radii—an effect not incorporated in our current model. The challenge of including energy-dependent albedo is discussed in Section \ref{sec_limitation_refl_table}. All of these simplifying assumptions highlight areas where the current model can be improved to better reflect physical reality.

\section{Conclusion}
\label{sec_conclusion}
In this work, we further upgraded the existing MONK code to model reflection by treating the disk as a semi-infinite electron atmosphere. The portion of the incident flux that is not reflected is absorbed by the disk, thereby increasing its temperature above the purely Keplerian value. The mutual feedback between the extended corona and the disk enabled us to determine an equilibrium solution for a given disk–corona system within a fully general relativistic framework. Our key findings are summarized below:
\begin{itemize}    
    \item A static slab corona covering the disk cannot account for the hard state observations of XRBs, particularly for photon indices $\Gamma < 1.7$, consistent with earlier studies based on local disk–corona equilibrium.
    \item For a slab coronal geometry in equilibrium with the disk, higher black hole spin and higher electron temperature lead to harder spectra; however, they do not significantly reduce $\Gamma_{\rm min}$, the lowest photon index achievable from a disk–corona system in equilibrium.
    \item As in previous studies, $\Gamma_{\rm min}$ can be significantly reduced by increasing the disk albedo (i.e., its ionization); however, this requires fine-tuning to remain consistent with observations (see Section \ref{sec_slab_corona_hard_state}).
    \item Although the disk and corona are in global equilibrium, achieving local balance at each radius requires either an outward flow of energy within the corona or a radial decrease in coronal temperature to reduce the energy transferred through Comptonization (see Section \ref{sec_local_global_balance}).
    \item Further work is needed to model reflection more accurately, including detailed atomic transitions. Other plausible geometries, such as a truncated disk or an outflowing corona, have the potential to produce harder spectra and will be explored in future studies.
\end{itemize}

\begin{acknowledgements}
SRD would like to thank Roberto Taverna and Andrzej Zdziarski for the discussions about reflection and Comptonization. SRD, MB, MD thank GACR project 21-06825X for the support, as well as the institutional support from RVO:67985815. WZ  acknowledges NSFC grant 12333004, and support by the Strategic Priority Research Program of the Chinese Academy of Sciences, grant no. XDB0550200.\end{acknowledgements}

\bibliographystyle{aa} 
\bibliography{ref} 

\begin{appendix}

\section{Theoretical estimations of luminosities}
\label{appendix_theoretical_estimates}
If the disk-corona system is in a global equilibrium then the disk and coronal luminosities do not remain independent any more. Here we present the estimation of luminosities expected from different components when the disk-corona is in equilibrium, very similar to what presented in \cite{Haardt1991}, HM93. We assume that the $\alpha$ fraction of the available accretion power ($L_{\rm acc}$) is available for intrinsic disk dissipation (blackbody radiation) and the rest (1-$\alpha$) is dissipated in the corona to re-energize hot electrons that scatter the photons. Therefore,
\begin{equation}
    L_{\rm intr}=\alpha L_{\rm acc}, L_{\rm sc}=(1-\alpha)L_{\rm acc}.
\end{equation}
Further, the albedo fraction of the illuminated luminosity reflected back from the disk gives 
\begin{equation}
    L_{\rm refl}={\rm albedo}\times L_{\rm ill}.
\end{equation}
Here we assume the anisotropy fraction $f_{\rm a}$ of the coronal luminosity goes downward and illuminates the disk ($\eta$ in HM93). For isotropic coronal radiation, $f_{\rm a}=0.5$. In addition, we assume the escaping fraction through corona is $f_{\rm e}$. Therefore, $f_{\rm e}$ fraction of the luminosity from the disk (blackbody emission + reflection) which enters the corona from below escapes through, and rest contributes to the total coronal luminosity. Therefore, total coronal luminosity ($L_{\rm corona}$) and the illumination to the disk becomes
\begin{equation}
    L_{\rm corona}=(1-f_{\rm e})(L_{\rm disk}+L_{\rm refl})+L_{\rm sc},
\label{eqn_L_corona}
\end{equation}
and
\begin{equation}
L_{\rm ill}=f_{\rm a} L_{\rm corona}.
\end{equation}
The above steps finally lead to the luminosity emitted by the disk, which comes from intrinsic dissipation as well as absorption of illumination:
\begin{multline}
    L_{\rm disk}=\alpha L_{\rm acc}+(1-{\rm albedo})L_{\rm ill},\\
    \implies L_{\rm disk}=\frac{\alpha+f_{\rm a}-\alpha f_{\rm a}-{\rm albedo}(f_{\rm a}-\alpha f_{\rm a}f_{\rm e})}{1-f_{\rm a}+f_{\rm a}f_{\rm e}}L_{\rm acc}.
\label{eqn_disk_emission}
\end{multline}
As the disk is completely covered by the slab corona for our case, the observed disk luminosity is
\begin{equation}
        L_{\rm disk, obs}=f_{\rm e}\,L_{\rm disk}.
\label{eqn_lum_disk_obs}
\end{equation}
Similarly, we can reduce the expressions of luminosity from other components as well in terms of $\alpha$, $f_{\rm e}$, $f_{\rm a}$ and albedo. Summing up all the luminosity components, the total luminosity reaching to the observer at infinity from the disk-corona system is
\begin{equation}
\label{eqn_Ltot_analytic}
    L_{\rm tot}=f_{\rm e}(L_{\rm disk}+L_{\rm refl})+(1-f_{\rm a})L_{\rm corona}=L_{\rm acc},
\end{equation}
which is the basic assumption for the total disk-corona system to be in global energy balance. For a given disk-corona system (including specifying the albedo) equilibrium between disk and corona provides unique values of $\alpha$, $f_{\rm e}$ and $f_{\rm a}$. We note that at equilibrium all these three parameters are coupled. Using these values, we can estimate the observed disk luminosity using equation \ref{eqn_lum_disk_obs} corresponding to the given disk-corona system. Similarly we also can estimate the observed luminosity from corona as well as due to reflection. We note that the $L_{\rm tot}$ of analytic estimation in equation \ref{eqn_Ltot_analytic} is the same as in equation \ref{eqn_total_luminosity_monk} with the difference of only the luminosity components contributed by photons from self-irradiation or lost in the BH as these components come in the MONK computations solely due to general relativistic effects which is ignored in the simple analytic estimation.

The numerical scheme of achieving equilibrium solution by iterating $\alpha$ (as described in Sect. \ref{sec_procedure_steps}) fixes the $f_{\rm e}$, $f_{\rm a}$ self-consistently. In the limit of very very optically thin corona ($\tau\rightarrow0$), equilibrium solution gives $\alpha\rightarrow1$ (Fig. \ref{fig_alpha_flux_fraction}(a)), $f_{\rm e}\rightarrow1$ (Fig. \ref{fig_leak_aniso}). Putting these values in equation \ref{eqn_L_corona}-\ref{eqn_lum_disk_obs} we find that irrespective of actual value of $f_{\rm a}$ and albedo, $L_{\rm corona}=L_{\rm sc}=0$ and $L_{\rm disk,obs}=L_{\rm disk}=L_{\rm acc}$. This gives the expected result that if the corona ceases to exist, the total accretion power is dissipated in the disk and we observe only the disk component.

The analytical estimations presented above is very similar to what presented in HM93 to find the equilibrium solution. However, although the assumption of disk luminosity is the same in both the works (equation 2a in HM93), there is a fundamental difference about the assumption of coronal luminosity. Compared to HM93, we introduced the parameter $f_{\rm e}$ and assume the $L_{\rm corona}$ to be the sum of the available accretion power in the corona and the luminosity from the disk interacting with the hot electrons (equation \ref{eqn_L_corona}). On the other hand, HM93 assumed that the coronal luminosity is only the accretion power available in the corona, $L_{\rm corona,HM}=L_{\rm sc}$ (equation 3b in HM93). We can mimic the convergence of our formalism with HM93 by assuming $f_{\rm e}=1$ without putting any constraint on $\alpha$.

Now, if we approximate $f_{\rm e}=1$, we find $L_{\rm corona}=L_{\rm sc}=(1-\alpha)L_{\rm acc}$, and $L_{\rm disk}=\left(\alpha+f_{\rm a}-\alpha f_{\rm a}-{\rm albedo}(f_{\rm a}-\alpha f_{\rm a})\right)L_{\rm acc}$. Therefore, with the limit of $f_{\rm e}=1$, the ratio between the two becomes 
\begin{multline}
    \frac{L_{\rm corona}}{L_{\rm disk}}=\frac{1-\alpha}{\alpha+f_{\rm a}-\alpha f_{\rm a}-{\rm albedo}(f_{\rm a}-\alpha f_{\rm a})}\\
    =\frac{1-\alpha}{1-(1-\alpha)+f_{\rm a}(1-\alpha)-{\rm albedo}\times f_{\rm a}(1-\alpha)}
\end{multline}
Following the notation used in HM93, if we replace (1-$\alpha$) by $f$, albedo by ${\rm a}$, and $f_{\rm a}$ by $\eta$, we get back the same equation (5a) in HM93.

It should be noted that in our formalism of equilibrium solution for a given disk-corona system, $f_{\rm e}$ is coupled with $\alpha$. Here to show the convergence of our formalism to HM93 by putting $f_{\rm e}=1$ without putting any constraint on $\alpha$ simply indicates that we assume the same as in HM93, $L_{\rm corona}$ consists only the accretion power available in the corona ($L_{\rm sc}$). Once we make our fundamental assumptions same, we reach to the same equations of luminosities of HM93 as expected.

\end{appendix}

\end{document}